\documentclass{phyeauth}
\usepackage{graphicx}
\usepackage{amsmath}
\usepackage{amssymb}

\begin{document}

\begin{frontmatter}
\title{Collective Transport in Bilayer Quantum Hall Systems}
\author[address1,address2]{Anton A. Burkov},
\author[address1,address3,address4]{Yogesh N. Joglekar},
\author[address1]{Enrico Rossi}, 
and
\author[address1]{Allan H. MacDonald}

\address[address1]{Physics Department, University of Texas at Austin, Austin TX 
78703, USA}
\address[address2]{Department of Physics. University of California at Santa 
Barbara, Santa Barbara CA 93106, USA}
\address[address3]{Department of Physics and Astronomy, University of Kentucky, 
Lexington KY 40506, USA}
\address[address4]{Theoretical Division, Los Alamos National Laboratory, Los Alamos 
NM 87544, USA} 

\begin{abstract}
Filling factor $\nu=1$ incompressible states in ideal bilayer quantum Hall systems 
have spontaneous interlayer phase coherence and can be regarded either as 
easy-plane pseudospin ferromagnets or as condensates of excitons formed from 
electrons in one layer and holes in the other layer. In this paper we discuss 
efforts to achieve an understanding of the two different types of transport 
measurements (which we refer to as drag and tunneling experiments respectively) 
that have been carried out in bilayer quantum Hall systems by the group of Jim 
Eisenstein at the California Institute of Technology. In a drag experiment, current 
is sent through one of the two-layers and the voltage drop is measured in the other 
layer. We will argue that the finding of these experiments that the voltage drop in 
the drag layer is different from that in the the drive layer, is an experimental 
proof that these bilayers do not have quasi-long-range excitonic order. The 
property that at $\nu=1$ the longitudinal drag voltage increases from near zero 
when spontaneous coherence is initially established, then falls back toward zero as 
it becomes well established, can be explained as a competition between the broken 
symmetry and the gap to which it gives rise. In the tunneling experiment, current 
is injected in one layer and removed from the other layer. The absence of 
quasi-long-range order likely explains the relatively small tunneling conductance 
per area found in the these measurements.
\end{abstract} 

\end{frontmatter}


\section{Introduction}
Among the broken symmetry states that occur in many-particle systems, those in 
which long range phase coherence is established, either for bosons \cite{becref} or 
for pairs of fermions \cite{bcstheory}, have special significance because of the 
quantum nature of their macroscopic order and because of the sometimes startling 
phenomenology. In semiconductors, the possibility of long-range phase coherence 
due to Bose condensation of electron-hole bound states (excitons) was first raised 
\cite{keldysh} nearly 40 years ago. The physics of excitonic Bose condensation is 
interesting in bilayer systems in which excitons can form from electrons in one 
layer and holes in the neighboring layer. This case is especially exciting because 
the possibility of making separate electrical contact to the two layers enables 
novel probes of superfluid transport phenomena. In the quantum Hall regime, because 
of Landau level degeneracy, magnetoexcitons can emerge from electrons and holes 
that both originate from the conduction band, vastly simplifying the task of 
realizing the high-density electron-hole fluids in which these phenomena are 
expected to be most robust. Excitonic Bose condensation in this case, it turns out, 
is expected to occur when the total Landau level filling factor of the bilayer 
system is near $\nu=1$. Although the anomalous transport properties discovered in 
bilayer quantum Hall systems \cite{jpeexpt} near this filling factor are thought 
\cite{bilayertheory,firsttheory} to follow from excitonic condensation and 
spontaneous phase coherence, it has not yet been possible to provide a complete 
interpretation of the observations.

In the present paper we concentrate on drag transport experiments, in which current 
flows in only one layer, but voltages are measured in both layers. The key 
observation in these experiments is that the longitudinal and Hall voltages 
measured in the current carrying layer and the electrically open layer are similar. 
This property suggests that the current is carried by quasiparticles that have 
weight in both layers even though interlayer tunneling amplitudes are negligibly 
small. It is naturally accounted for by a BCS-like mean-field theory of bilayer 
excitonic Bose condensation in which quasiparticles that are a coherent combination 
of states localized in separate layers are analogous to the coherent 
electron-plus-hole quasiparticles of superconductors; the indefinite layer index of 
these quasiparticles follows from the state's broken symmetry in the same way as 
the indefinite charge of BCS quasiparticles follows from superconducting order. In 
this picture, because the quasiparticles carry current that is divided equally 
between the layers, the drag experiment constraint that net current flow only 
through one layer forces an excitonic supercurrent through the bilayer. However, as 
we explain in greater detail below, the experimental observation that the voltages 
measured in the two layers are similar but not identical implies \cite{vignale} that 
this collective superflow is not completely dissipationless. In the phenomenology 
that we explain below, dissipation is accounted for by the flow of vortices in the 
order-parameter field. 


\section{Phenomenology} 
In bilayers dc transport is characterized by a $4 \times 4$ conductivity tensor 
since its labels have both layer and two-dimensional Cartesian indices. For 
balanced layers, invariance under interchange of layer indices guarantees that even 
($+$) and odd ($-$) channel response functions decouple. For isotropic layers these 
$2 \times 2$ tensors are characterized by their Hall and longitudinal 
conductivities so that there are four independent linear response coefficients:
\begin{eqnarray}
j_{\pm,x} &=& \sigma_{\pm} \frac{E_{T,x} \pm E_{B,x}}{2} 
+\sigma^{H}_{\pm} \frac{E_{T,y} \pm E_{B,y}}{2} \nonumber \\
j_{\pm,y} &=& - \sigma^{H}_{\pm} \frac{E_{T,x} \pm E_{B,x}}{2}
+\sigma_{\pm} \frac{E_{T,y} \pm E_{B,y}}{2}.
\label{notations}
\end{eqnarray} 
where $j_{\pm,\alpha}\equiv j_{T,\alpha} \pm j_{B,\alpha}$ and $T$ and $B$ label 
top and bottom layers respectively. These relations can be inverted to find the 
four corresponding independent resistivity coefficients, $\rho_{\pm}$ and 
$\rho^{H}_{\pm}$. In the absence of excitonic condensation, apart from the weak 
inter-layer scattering processes that give rise to small drag voltages under normal 
circumstances and are neglected here, currents in a layer produce an electric field 
only in the same layer implying that $\rho_{+} = \rho_{-}$ and 
$\rho^{H}_{+} = \rho^{H}_{-}$. When an excitonic Bose condensate with quasi 
long-range order is present, {\it i.e.} at temperatures below the 
Kosterlitz-Thouless temperature, the odd (-) channel linear response Hall and 
longitudinal resistivities should vanish because \cite{vignale} any difference in 
electric field between the layers would be shorted out by electron-hole pair 
condensate superflow. Experiments demonstrate that the even (+) and odd (-) channel 
resistivities in bilayers differ dramatically only for closely spaced layers and 
only for $\nu \approx 1$, thus strongly supporting the belief that collective 
excitonic transport is occurring in these systems. However, experiments also show 
that {\it the odd channel resistivities do not vanish}. It is these finite but 
non-zero odd channel resistivities that we concentrate on in this paper, since we 
believe that they are very revealing probes of the order that occurs in the system.

Weakly resistive transport in two-dimensional superfluids is normally understood in 
terms of Magnus-force driven vortex motion. Vortex flow leads to a steady rate of 
change of interlayer phase that differs at different points in the sample, and 
therefore via the Josephson relationship, gives rise to an odd (-) channel electric 
field. There are two reasons we expect vortex-motion induced dissipation to be 
significant in quantum Hall bilayers. First of all, vortices in $\nu=1$ quantum 
Hall superfluid carry electrical charge $e^{*}=e/2$, and the finite-energy 
integer-charge elementary excitations of quantum Hall bilayers can be thought of, 
at least approximately, as being composed of bound vortex pairs 
\cite{yangmoon,breycharge} with opposite vorticity. When the filling factor of a 
quantum Hall bilayer deviates from $\nu=1$, many charges of this nature are 
nucleated in the incompressible state background. Even at $\nu=1$, long-length 
scale inhomogeneity in the system will nucleate charged quasiparticles. Measurement 
of a finite odd channel linear resistivities suggests that in real samples some of 
these vortices are free even in the absence of the Magnus force associated with pair 
condensate currents; free vortices will always lead to voltages linear in current. 
The second unique feature of quantum Hall superfluids which opens up an opportunity 
for vortex transport to play an important role is that the quantum Hall effect 
causes both even (+) and odd (-) channel longitudinal quasiparticle resistivities to 
vanish in the limit of zero temperature, even when there is no collective 
transport. In effect, in quantum Hall ferromagnets, we are able to look at 
vortex-flow dissipation in a conductor which is nearly dissipation-free even in the 
normal state.


\section{SCBA Calculations} 
We argue below that it is possible to separate quasiparticle and condensate 
contributions to transport coefficients in these systems. This argument is based 
partly on microscopic self-consistent Born approximation calculations for 
disordered quantum Hall superfluids that we now discuss. The point of view taken 
below in assessing the experimental results is based primarily on the type of 
result presented in this section, which, in turn, is based on an approximation for 
charge and pseudospin response in quantum Hall superfluids explained in detail in 
previous work \cite{joglekarfluc,joglekardcj}. The discussion presented here is 
purely qualitative. These calculations treat interactions via a generalized RPA 
approximation and disorder via a self-consistent Born approximation. In this 
treatment, excitonic superfluidity (incorrectly \cite{firsttheory}) occurs at any 
layer separation $d$ no matter how large, but the phase transition between ordered 
and disordered states can be (correctly) driven by increasing the degree of 
disorder in the system. 
\begin{figure}[h]
\begin{center}\leavevmode
\includegraphics[width=0.7\linewidth]{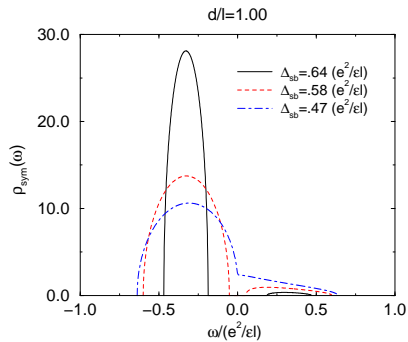}
\caption{Symmetric quasiparticle density-of-states, $\rho_{sym}(\omega)$ for a 
bilayer system with $\nu=1$ and different degrees of disorder, characterized by 
different values of $\Delta_{sb}$. $\Delta_{sb}$ is the exchange self-energy that 
favors symmetric quasiparticles over antisymmetric quasiparticles when the 
coherence phase angle is set to zero, a quantity that is proportional to the order 
parameter in SCBA+HF theory. For $\nu=1$, the antisymmetric quasiparticle 
density-of-states satisfies $\rho_{asym}(-\omega)=\rho_{sym}(\omega)$. The odd (-) 
channel conductivity is finite only when both densities of states are finite at the 
Fermi energy $\omega=0$.}
\label{fig:one}
\end{center}
\end{figure} 
In the ordered state, the occupied quasiparticle states spontaneously develop 
interlayer phase coherence. For coherence angle equal to zero and balanced 
bilayers, the quasiparticles experience an anomalous self-energy of collective 
origin that acts like a strong interlayer tunneling term (real with amplitude 
$\Delta_{sb}/2$) which establishes the interlayer coherence, in addition to the 
random potentials that exist in each layer. Without disorder the Landau level 
density-of-states $\rho(\omega)$ of the quasiparticles would consist of two delta 
function pieces, the symmetric and antisymmetric branches separated in energy by 
$\Delta_{sb}$. The ordering energy competes with the random potential by broadening 
the Landau levels and limiting the extent to which the quasiparticles can take 
advantage of a difference in potential between the layers. The densities of states 
in Fig.~\ref{fig:one} are plotted for three different disorder strengths, or 
equivalently, three different order parameter values. In the presence of disorder 
all quasiparticle states have greater weight in one layer than in the other and 
have mixed character when projected onto symmetric and antisymmetric states. The 
order parameter of the bilayer can be defined as the difference per Landau level 
orbital of quasiparticle symmetric and antisymmetric projections summed over all 
occupied levels. By this measure, the order parameter at $\nu=1$ approaches one as 
the disorder in the system weakens. In the SCBA, an artificial gap in the 
quasiparticle density of states arises when the disorder is sufficiently weak; in a 
more realistic calculation the quasiparticle density of states at the Fermi energy 
would decline continuously with the strength of the order but never vanish.

\begin{figure}[h]
\begin{center}\leavevmode
\includegraphics[width=0.7\linewidth,angle=0]{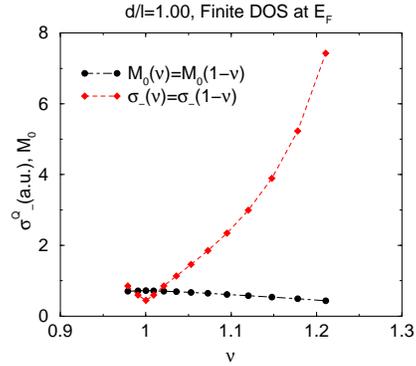}
\caption{Odd (-) channel conductivity (in arbitrary units) and order parameter as a 
function of filling factor for a fixed random potential strength. The conductivity 
and the order parameter are symmetric around $\nu=1$. These results follow from 
SCBA calculations of linear response functions for a model with a random disorder 
potential in each layer and no correlations between the potentials in the different 
layers. A finite odd (-) channel conductivity requires quasiparticles that can carry 
different currents in the two layers and is reduced as order develops because the 
degree of layer polarization of typical quasiparticles is proportional to 
$\Gamma/\Delta_{sb}$, where $\Gamma$ is the Landau level width.}
\label{fig:two}
\end{center}
\end{figure}

\begin{figure}[b]
\begin{center}\leavevmode
\includegraphics[width=0.7\linewidth,angle=0]{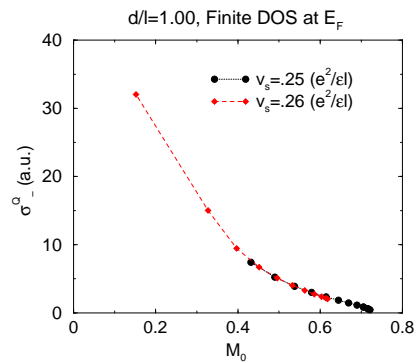}
\caption{Odd (-) channel conductivity (in arbitrary units) as a function of order 
parameter $M_0$. These results are calculated for various filling factors and two 
different disorder strengths. The approximate coincidence of these two curves 
demonstrates that the order parameter $M_0$ which depends on the same two 
variables, is the most important factor in determining the conductivity value.}
\label{fig:three}
\end{center}
\end{figure}

In the normal state, current is carried independently in the two layers, implying 
that the even (+) and odd (-) channel conductivities are identical; currents 
produce voltages only in the layer in which they flow. In the ordered state, the 
strong effective tunneling amplitude leads to quasiparticle states that tend to 
have their charge evenly divided between the two layers. These quasiparticle states 
tend to carry currents that are also nearly equally divided between the two layers, 
causing the odd (-) channel conductivity to be much smaller than its even (+) 
channel counterpart.  The odd (-) channel conductivity and the order parameter are 
plotted as a function of filling factor in Fig.~\ref{fig:two}; we see here that the 
odd channel conductivity is strongly suppressed near $\nu=1$ where the spontaneous 
coherence is most well developed. In Fig.~\ref{fig:three} we plot the odd (-) 
channel conductivity as a function of the order parameter for two different 
disorder strengths and various filling factors, demonstrating that the suppression 
is more strongly connected to the degree of order than it is to either the filling 
factor or the disorder potential strength. Since the density of states at Fermi 
energy is strongly suppressed near $\nu=1$, both conductivities are actually 
reduced in the ordered state, but the odd channel conductivity is reduced much more 
significantly.

Because we expect (and know from experiment) that Hall angles are very large in the 
quantum Hall regime, the longitudinal resistivities in each channel should be 
nearly proportional to the longitudinal conductivities. The difference between even 
(+) and odd (-) channel conductivities explained above should therefore lead to the 
same relative magnitudes for even (+) and odd (-) channel longitudinal 
resistivities, and therefore to electric fields that have the same sign in drive 
and drag layers. This is opposite to what is observed experimentally. We believe 
that the experimental observations can be explained only by positing a condensate 
contribution to the currents carried through the system.


\section{Condensate Conductivity} 
Electron-hole pairs can carry opposing currents in the two layers and therefore a 
condensate can contribute to the odd channel conductivity. However, any vortices 
that are not pinned, will flow in the presence of a condensate current and produce 
odd channel electric fields. These electric fields will also drive extended state 
quasiparticles to carry current as discussed in the previous section. When vortices 
are unpinned, their thermal and quantum fluctuations must lead to a loss of 
long-range phase coherence. The observation of odd channel electric fields 
therefore suggests that the bilayers that have been studied have unpinned vortices.

Since the quasiparticles and the condensate phase (through its Josephson relation) 
see the same electric field, it follows that their conductivities $\sigma^{Q}$ and 
$\sigma^{C}$ add, {\it i.e.} they can be regarded as two separate contributions 
that carry odd channel current in parallel:
\begin{eqnarray}
\sigma_{-} &=& \sigma^{Q}_{-} + \sigma^{C}_{-} \nonumber \\\
\sigma^{H}_{-} &=& \sigma^{QH}_{-} + \sigma^{CH}_{-} \nonumber \\
\label{separation}
\end{eqnarray} 
The discussion in the previous section allowed only for $\sigma^{Q}_{-}$. The 
observed sign of the longitudinal Hall voltage implies that 
$\sigma_{-} \gg \sigma_{+}$, and since we have argued that $\sigma^{Q}_{-}$ must be 
smaller than $\sigma_{+}$, this implies that $\sigma_{-} \approx\sigma^{C}_{-}$. 
Measuring the odd channel longitudinal conductivity should therefore provide a 
direct measurement of condensate current in the bilayer.

As emphasized above, we do not believe that it is possible to interpret the 
experiments without positing such a collective current. $\sigma^{Q}_{-}$ is small 
when order is well developed partly because the quasiparticles of the bilayer in an 
ordered system experience an in-plane pseudospin field that reduces the degree of 
layer polarization due to fluctuations in the local difference between random 
potentials in the two layers. Because the quasiparticles have little layer 
polarization even in the presence of disorder $\sigma^{Q}_{-}/\sigma_{+}$ is 
$\sim (\Gamma/\Delta_{qp})^2$ where $\Gamma$ is the Landau level width and 
$\Delta_{qp}$ is the mean-field charge gap. 

The quantum Hall effect occurs in the even (+) channel, not in the odd (-) channel. 
We should therefore expect that the odd (-) channel Hall conductivity vanishes in 
the limit of zero temperature if both quasiparticles and vortices are localized in 
this limit. Indeed, this property is already suggested by current experiments, 
since the Hall voltages measured in drive and drag layers seem to approach each 
other at very low temperatures. From the SCBA linear response theory for the 
quasiparticle conductivities, it is clear that for $\nu \approx 1$ the 
quasiparticle Hall angles will be similar and large in both even and odd channels, 
{\it i.e.} that $\sigma_{+}^{H} \gg \sigma_{+}$ and 
$\sigma_{-}^{QH} \gg \sigma_{-}^{Q}$. We do not, however, have a clear idea at 
present of the Hall angle for the condensate conductivity, which is related to the 
relationship between vortex flow and condensate current directions. We believe that 
further experiments, analyzed with the picture explained here, should allow this 
subtle issue, long controversial \cite{vortexhall} in the case of superconductors, 
to be settled for the case of quantum Hall ferromagnets.


\section{Other Open Issues}
We have seen that the drag experiments in bilayer quantum Hall systems 
simultaneously provide strong evidence that collective transport by an 
electron-hole pair condensate occurs in these systems, and that it is accompanied 
by dissipation suggesting that unpinned vortices are always present. The presence 
of free vortices may help explain the surprisingly small interlayer conductance 
that can be inferred from the tunneling experiments. In our view the well developed 
even (+) channel quantum Hall effect that is seen in these systems is evidence of a 
large {\it local} order parameter. If there were long range order, this property 
would\cite{icps26} be difficult to reconcile with the experimentally measured layer 
to layer conductance per electron which is many orders of magnitude smaller than 
$e^2/h$. When current is injected on one side of one layer and extracted from the 
opposite side of the other layer, the overall resistance is still limited by weak 
hopping between the layers. It appears that collective tunneling of electrons 
between the layers is strongly suppressed \cite{firsttheory}. 
Experiments \cite{jpeexpt} can give us results for the temperature-dependent 
vortex-flow resistivity, do give us results for the height and width of the 
low-bias peak in the tunneling conductance, and do give us results for the in-plane 
field scale at which the tunneling I-V characteristic changes character. The 
challenge for theory is to provide a common explanation for all these phenomena 
which, it appears, must have a common origin. 

This work was supported by the Welch Foundation and by the NSF under grant 
DMR-0115947 at the University of Texas, and by the NSF under grant DMR-0071611 at 
the University of Kentucky.



\end{document}